\documentclass[universe,article,accept,pdftex,moreauthors]{mdpi} 
\firstpage{1} 
\pubvolume{1}
\issuenum{1}
\articlenumber{0}
\pubyear{2025}
\copyrightyear{2025}
\externaleditor{Firstname Lastname} % More than 1 editor, please add `` and '' before the last editor name
\datereceived{6 March 2025} 
\daterevised{11 April 2025} % Comment out if no revised date
\dateaccepted{15 April 2025} 
\datepublished{ } 
\hreflink{https://doi.org/} % If needed use \linebreak
\newcommand{\rrab} {\mbox{RR\emph{ab}}}
\newcommand{\rrc} {\mbox{RR\emph{c}}}

\usepackage{mathtools}
\usepackage{amsmath}
\usepackage{adjustbox}
\usepackage{multirow}
\usepackage{upgreek}
\usepackage{amssymb}
\usepackage{graphicx}
\usepackage{graphics}
\usepackage{multirow}
\usepackage{color}
\usepackage{xcolor}
\usepackage{bm}	
\usepackage{graphicx}	% Including figure files
\usepackage{amssymb}

\usepackage{comment}
\usepackage{upgreek}
\usepackage{float}
\usepackage{graphicx}

\def\apj{Astrophys. J.}

\def\apjl{ Astrophys. J. Lett.}

\def\aap{Astron. Astrophys.}
\def\mnras{Mon. Not. R. Astron. Soc.}
\def\pasp{PASP}

\def\nat{Nature}
\def\aj{Astron. J.}

\Title{Determining the Scale Length and Height of the Milky Way's Thick Disc Using RR~Lyrae%attention: title altered
}

\TitleCitation{Determining the Scale Length and Height of the Milky Way's Thick Disc Using RR~Lyrae}

\Author{
Roman 
 Tkachenko $^{1}$\orcidA{}, 
Katherine Vieira $^{2,}$*\orcidB{}, 
Artem Lutsenko $^{3,4}$\orcidC{}, 
Vladimir Korchagin $^{1}$\orcidD{} \linebreak  and 
Giovanni Carraro $ ^{3}$\orcidF{}}

\AuthorNames{Roman Tkachenko, Katherine Vieira, Artem Lutsenko, Vladimir Korchagin, and Giovanni Carraro}

\isAPAStyle{%
       \AuthorCitation{Tkachenko, R., Vieira, K., Lutsenko, A., Korchagin V.\& Carraro, G.}
         }{%
        \isChicagoStyle{%
        \AuthorCitation{Lastname, Firstname, Firstname Lastname, and Firstname Lastname.}
        }{
        \AuthorCitation{Tkachenko, R.; Vieira, K.; Lutsenko, A.; Korchagin, V.; Carraro, G.}
        }
}

\address{%
$^{1}$ \quad Institute of Physics, Southern Federal University, Stachki 194, Rostov-on-Don 344090, Russia; rtkachenko@sfedu.ru (R.T.); vkorchagin@sfedu.ru (V.K.) \\
$^{2}$ \quad Instituto de Astronom\'ia y Ciencias Planetarias, Universidad de Atacama, Copayapu 485, \linebreak  Copiap\'o 1531772, Chile\\
$^{3}$ \quad Dipartimento di Fisica e Astronomia, Universitá di Padova, Vicolo Osservatorio 3, 35122 
Padova, Italy; artem.lutsenko@studenti.unipd.it (A.L.); giovanni.carraro@unipd.it (G.C.)  \\
$^{4}$ \quad INAF-Padova Observatory, Vicolo dell’Osservatorio 5, 35122 Padova, Italy}

\corres{Correspondence: katherine.vieira@uda.cl}

\abstract{Using the RR Lyrae surveys Gaia DR3 Specific Objects Study,  PanSTARRS1 and ASAS-SN-II, we determine the Milky Way's thick disc scale length and scale height as well as the radial scale length of the galaxy's inner halo.  
We use a Bayesian approach to estimate these values using two independent techniques: Markov chain Monte Carlo sampling, and importance nested sampling. We consider two vertical density profiles for the thick disc. In the exponential model, the scale length of the thick disc is $h_R=2.14_{-0.17}^{+0.19}$ kpc, and its scale height is  $h_z=0.64_{-0.06}^{+0.06}$ kpc. In the squared hyperbolic secant profile $sech^2$, those values are correspondingly $h_R=2.10_{-0.17}^{+0.19}$ kpc and $h_z=1.02_{-0.08}^{+0.09}$ kpc. The density distribution of the inner halo can be described as a power law function with the exponent $n =-2.35_{-0.05}^{+0.05}$ and flattening $q =0.57_{-0.02}^{+0.02}$. We also estimate the halo to disc concentration ratio as $\gamma=0.19_{-0.02}^{+0.02}$ for the exponential disc and $\gamma=0.32_{-0.03}^{+0.03}$ for the $sech^2$ disc.}

\begin{document}

%%%%%%%%%%%%%%%%%%%%%%%%%%%%%%%%%%%%%%%%%%

\section{Introduction}
The study of galactic structures is one of the most important topics of modern astrophysics since it provides insights into the formation of the Milky Way galaxy. Among~all the galactic components, the~precise structure of the thick disc of the Milky Way galaxy plays an important role in influencing the distribution and equilibrium rotation of the Milky Way's thin and thick discs, as well as in inferring the dark matter density distribution in the solar neighborhood and inner halo. Thick discs were first discovered in external galaxies by \citet{Burstain}. The~existence of the thick disc in the Milky Way galaxy was confirmed immediately after by \citet{GR}. 

The presence of the galactic thick disc can be confirmed not only by the vertical density distribution of the disc. Kinematically, the~rotation of the thick disc lags behind that of the thin disc by about 50 km/s~\cite{Bensby11}. The~thick disc is also older, metal-poorer and richer in alpha elements compared to the thin disc~\cite{WG,Bensby7,Fuh,Bensby11,Reddy}. The~knowledge of the radial and vertical density distributions in the thick disc is important for understanding star formation history in the Milky Way galaxy, given that they dictate stellar migration and play a role in the overall dynamics of the inner galaxy. Several attempts have been made to determine the parameters of the thick disc of the Milky Way using different stellar populations. Estimated values of the thick disc scale length vary from 1.9 to 4.7 kpc, and~the scale height from 0.51 to 1.36 kpc~\cite{R96,R14,Juric,Bensby11,BR,LH,Carollo10,MV, Vieira}. 

RR Lyrae (RRLs), being old objects of the Milky Way galaxy with typical ages exceeding 9--10 Gyr
~\cite{clemen,Garcia,Smith}, are metal-poor periodic pulsating variable stars located on the instability strip of the horizontal branch of the Herztzsprung--Russel diagram~\cite{Garcia}. They serve as excellent distance indicators due to a well-defined luminosity--period--metallicity relation~\cite{Murav}. The regular pulsation patterns of these stars and their relative abundance among older stellar populations make them valuable for probing the structure of the Milky Way's subsystems, particularly the thick disc and the halo. RRLs have been used in various contexts to study the structure and kinematics of the Milky Way galaxy, contributing significantly to our understanding of properties of the Milky Way's subsystems, particularly the thick disc~\cite{MV,Layden,Kinemuchi,MVP} and the halo~\cite{MV,Sesar13,Sesar11,wat,Zinn,TK}. 

Due to the complex interplay between various galactic components and poor statistics,  uncertainties remain regarding the parameters of the Milky Way's thick disc. The~Gaia mission opened a new possibility in studying the properties of the Milky Way's thick disc by increasing the number of RR Lyrae stars with measured parameters by more than two orders of magnitude compared to previous studies. Based on the new survey data on RR Lyrae stars from the Gaia DR3 catalog, we re-evaluate the parameters of the Milky Way's thick disc. We employ statistical Bayesian techniques to determine the parameters of the thick disc, namely~Markov chain Monte Carlo (MCMC) sampling (as implemented by the 
 \texttt{emcee} \cite{Fore} Python package), 
 and~importance nested sampling (INS) (as implemented by the \texttt{Nautilus} \cite{nautilus} Python package). This allows us to derive more accurate parameters of the Milky Way's thick disc and halo compared to previous~studies.

The layout of this paper is as follows: Section~\ref{sec:obs} gives details of the observational data on the RRL sample on which we base our study and explains how we build the selection function necessary for the fitting procedure of the thick disc's parameters; Section~\ref{sec:model} presents the models of the galactic thick disc and halo and  explains the Bayesian approach used in the paper. Section~\ref{sec:res} presents the obtained values for the radial and vertical scale lengths of the thick disc together with the parameters of the halo. Section~\ref{sec:discuss} compares and discusses the derived results with previous studies. Section~\ref{sec:conclusion} summarizes the results of our~study.

%%%%%%%%%%%%%%%%%%%%%%%%%%%%%%%%%%%%%%%%%%
\section{Observational~Data }\label{sec:obs}
\unskip

\subsection{The~Sample}\label{sec:dats}

We base our study on the publicly available catalog of RR Lyrae stars: the Gaia DR3 Specific Objects Study~\cite{clemen} (G < 20.7 mag), ASAS-SN-II~\cite{ASAS} (G < 17 mag) and PanSTARRS1~\cite{PS1} (G < 21 mag). From~now on, we will refer to them as Gaia SOS, ASAS and PS1, respectively. Gaia SOS and ASAS are both all-sky, while PS1 covers about 3/4 of it. We cross-matched the Gaia SOS and ASAS with a 5$''$ tolerance, and~Gaia SOS and PS1 with a 3$''$ tolerance parameter, as~in \citet{Mat2020,ASAS2}. Then, as~in~\cite{MV,Mat2020}, we select only \rrab~type Lyrae stars and exclude \rrc~Lyraes because the latter have problems with completeness and contamination because the RRc contamination is significantly higher than the RRab. As~a result, we have 175,350 \rrab~Lyraes from Gaia SOS, ~61,829 stars from ASAS+PS1 combined and 53,424 cross-matched stars between the two samples, Gaia SOS and ASAS+PS1. 

\begin{figure}[H]
%\isPreprints{\centering}{} % Only used for preprints
\scalebox{.99}[0.99]{\includegraphics[width=14 cm]{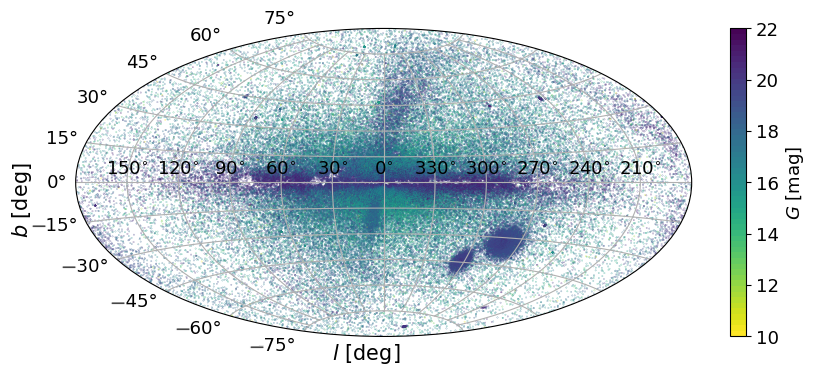}}
\caption{ Sky distribution   
 of the uniquely selected 183,755 \rrab~Lyraes in galactic coordinates, color-coded by their G-band magnitude. For~the 8495 stars in ASAS+PS1 but not in Gaia, a~G-band magnitude is obtained from their available photometry, following~\cite{Mat2020}.\label{fig1}}
\end{figure}

To convert the heliocentric coordinates into galactocentric ones, we use the \texttt{astropy} package \citep{astro1,astro2,astro3}, adopting the solar galactocentric distance $R_{\odot}=8.122$ kpc~\cite{Gravity} and the vertical displacement of the Sun from the galactic plane $Z_{\odot}=20.8$ pc~\cite{Zsun}.
Figure~\ref{fig1} shows the sky distribution in galactic coordinates of the above uniquely selected 183,755 \rrab~Lyraes, which reflects the main structures reported in other studies: the Milky Way's disc and halo, the~Magellanic Clouds, and~the Sagittarius Core and~Stream.

\unskip

\subsection{Selection~Function}\label{sec:SF}
Constructing a selection function for a specific sample involves the usage of various criteria to measure or weight the selected sample of objects of interest against their whole true population. In~other words, it aims to relate the observed number of objects $N_{obs}$ to the actual number of stars $N_{true}$ as $S=N_{obs}/N_{true}$. We define the selection function as a product of the two catalogs%please check that intended meaning has been retained
:
\begin{eqnarray}
    S = S_{combined}\; S_{flag} . \label{e:Cdef} 
\end{eqnarray}

\textls[-25]{For the first function $S_{combined}$, we use a procedure suggested by \mbox{\citet{Ryb}}} for two independent catalogs and recently used in other studies~\cite{Mat2020,Mat2024,Cantat}. We choose Gaia SOS as the first catalog and ASAS+PS1 as the second one. The~probability that a given RRL is selected in the combined catalog, Gaia+(ASAS+PS1), is the sum of the individual probabilities minus the probability that it is present in both. Then, the selection function of the combined catalogs can be written as follows:
\begin{equation} \label{eq:scomb}
\begin{split}
S_{\mathrm{combined}}~=~& S_{\mathrm{Gaia}} 
                           +  S_{\mathrm{ASAS+PS1}} \\
                        & - S_{\mathrm{Gaia}} \times  S_{\mathrm{ASAS+PS1}}  . 
\end{split}
\end{equation}
The procedure used here can be described as follows (for details see~\cite{Mat2020}). 
The completeness $S_A$ of the catalog $A$ can be considered as the probability of detecting a star in this catalog---that is, the~ratio of the number of the observed stars in the survey $N_A$ divided by the true number of stars $N_{\mathrm{true}}$ :
\begin{eqnarray}
    S_A = P_{A} = \frac{N_{A}}{N_{\mathrm{true}}} \label{e:SA}.
\end{eqnarray}

Assuming that the two catalogs are independent, we can write the number $N_{A\cap B}$ of the stars common to two catalogs as follows:
\begin{equation}
N_{A\cap B} = P_{A\cap B}N_{\mathrm{true}} = P_A P_B N_{\mathrm{true}} =
\frac{N_A}{N_{\mathrm{true}}} \frac{N_B}{N_{\mathrm{true}}} N_{\mathrm{true}} = S_A N_B \Longrightarrow
S_A = \frac{N_{A\cap B}}{N_B}
\label{eq:inter}.
\end{equation}
Similarly, for catalog B $\displaystyle S_B=\frac{N_{A\cap B}}{N_A}$. It is also immediately apparent from Equation~(\ref{eq:inter}) that $S_{A\cap B}=S_A S_B$. In~other words, by~mutual comparison of each independent catalog with their intersection, we can assess their individual completeness. This procedure relies significantly on two assumed facts: no contamination in each catalog, and no mismatches between them. This procedure is employed to compute the completeness of ASAS and PS1 individually, of~their combination (ASAS+PS1), of~Gaia SOS and of their full combination (Gaia SOS+ASAS+PS1).

\begin{figure}[H]
\scalebox{.99}[0.99]{\includegraphics[width=6.9cm]{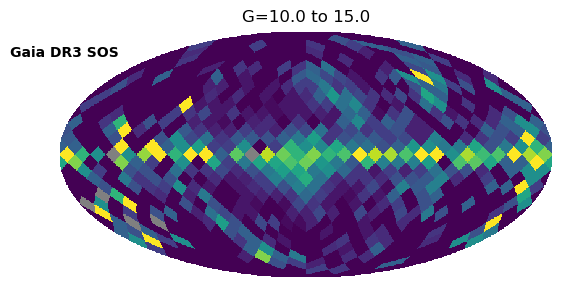}}
\scalebox{.99}[0.99]{\includegraphics[width=6.9cm]{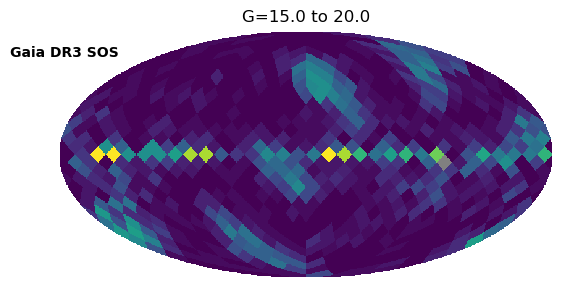}}
\scalebox{.99}[0.99]{\includegraphics[width=6.9cm]{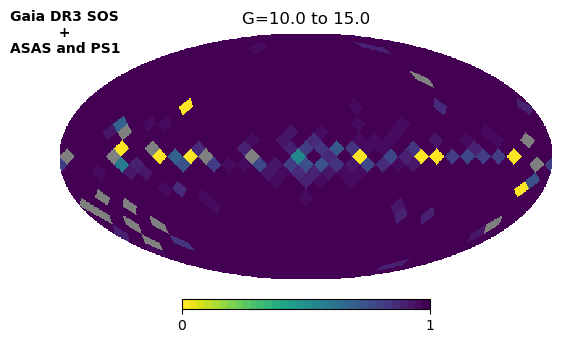}}
\scalebox{.99}[0.99]{\includegraphics[width=6.9cm]{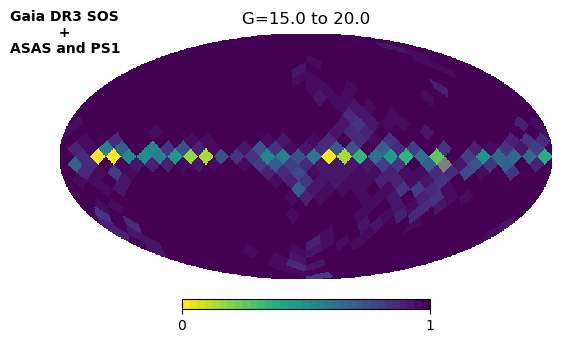}}
\caption{Completeness maps in galactic coordinates for the Gaia DR3 SOS (\textbf{top row}) and for the combined catalog $S_{combined}$ (\textbf{bottom row}) in two ranges of apparent G magnitude: G~=~10 to 15 (\textbf{left}) and G~=~15 to 20 (\textbf{right}).\label{figScomb}}
\end{figure}

Figure~\ref{figScomb} demonstrates the relative completeness of Gaia SOS (top row) and the completeness of the combined Gaia SOS+ASAS+PS1 catalog (bottom row) for two different intervals of apparent G magnitude.  In~this figure, we used the \texttt{healpy} Python software~\cite{heal,heal2} and split the sky into HEALPix regions of level 3, corresponding to approximately $7.4$ deg$^2$ in galactic coordinates. As~can be seen from the top row of Figure~\ref{figScomb}, there are patches of incompleteness in Gaia SOS which can be connected to the Gaia satellite scanning~law.

We then apply the second selection function, which discards stars with bad quality measurements according to several criteria. As~shown in previous works~\cite{Rimoldini19,clemen,GaiaVariable}, the~bad quality measurements are related to the crowded areas (galactic plane or bulge region) with a significant number of artifacts and spurious contaminants~\cite{Iorio}.  Following \citet{Garcia}, we consider the values of \texttt{ruwe} and \texttt{phot\_bp\_rp\_excess\_factor}, which are applied to Gaia SOS, and~$E(B-V)$, which is applied to both the Gaia SOS and ASAS+PS1 catalogs:
\begin{itemize}
    \item \texttt{ruwe}$>1.4$.
    \item \texttt{phot\_bp\_rp\_excess\_factor}$>1.5$
    \item $E(B-V)>0.8$
\end{itemize}{}

The above
 conditions characterize the stars that will be rejected. The~criterion imposed on \texttt{ruwe} (renormalized unit weight error) chooses sources whose astrometric solutions correspond well to a single-star five-parameter solution~\cite{Lindegreen18} and excludes blended sources or unresolved stellar binaries~\cite{Belokurov20}. The~\texttt{phot\_bp\_rp\_excess\_factor} gives the ratio between the combined flux in the Gaia $BP$ and $RP$ bands and the flux in the $G$ band. Large values of this ratio are caused by blended sources~\cite{EvansGaia}. The~third criterion, reddening $E(B-V)$, is applied to both Gaia SOS and ASAS+PS1~catalogs. 

\begin{figure}[H]
%\isPreprints{\centering}{} % Only used for preprints
\scalebox{.99}[0.99]{\includegraphics[width=14 cm]{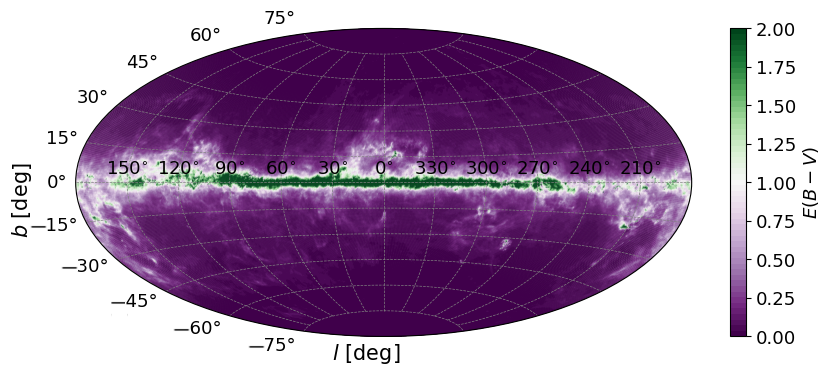}}
\scalebox{.99}[0.99]{\includegraphics[width=12 cm]{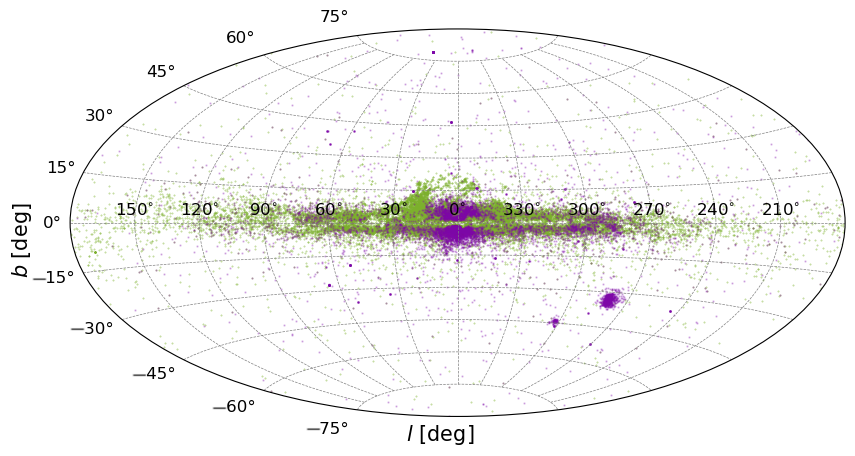}}
\caption{Top figure: Reddening map in galactic coordinates. Bottom figure: distribution of RRLs with ``bad'' \texttt{ruwe} $>1.4$ (purple points) and \texttt{phot\_bp\_rp\_excess\_factor} $>1.5$ (green points).\label{fig3}}
\end{figure}

Figure~\ref{fig3} shows the reddening map (top figure) taken from \citet{Shleg98} with 14\% correction taken from~\cite{Sch,Yuan}. The~bottom figure shows the distribution of the stars with a ``bad'' \texttt{ruwe} (purple points) and large BP/RP excess factor (green points).
As can be seen from Figure~\ref{fig3}, stars with bad quality measurements are concentrated near the galactic plane, in~the regions with high concentration of dust and blended sources. When determining the parameters of the galactic thick disc and halo, we will exclude from consideration the areas of the sky with latitudes $|b|<10^{\circ}$.

Let us introduce the selection function $S_{flag}$, which is connected to the quality cut, as~follows:
\begin{eqnarray}
    S_{flag} = N_{Good}/N_{Total} .\label{eq:Cflag} 
\end{eqnarray}
where $N_{Total}$ is the total number of stars and $N_{Good}$ is the number of stars not rejected by the three above-mentioned quality conditions. It should be noticed that if we have a star with a large \texttt{ruwe} or BP/RP excess factor in Gaia SOS but it is a reliable source in the ASAS+PS1 catalog, we select it as "good" and it is included in the $N_{Good}$ sample. We consider a star in ASAS+PS1 as reliable, following~\cite{Mat2024,Mat2020}\endnote{Data publicly available in the online archive at:  
\href{https://github.com/cmateu/rrl_completeness/tree/main/data/}{\tt https://github.com/cmateu/rrl\_completeness/tree/main/data/}.  This sample was used in \cite{Mat2024,Mat2020}.}. 
In the ASAS catalog, RRLs with periods $< 0.95$ d were discarded (to avoid the excess of spuriously identified stars) as well as those with suspiciously large amplitudes > 2 mag~\cite{Mat2020}. In~the PS1 catalog, 'bona fide' sources have a classification score $score_{3,ab}>0.8$, which, according to~\cite{PS1}, yields a sample with 0.97 purity and 0.92 completeness~\cite{Mat2020}. 

\begin{figure}[H]
%\isPreprints{\centering}{} % Only used for preprints
\scalebox{.98}[0.98]{\includegraphics[width=6.9 cm]{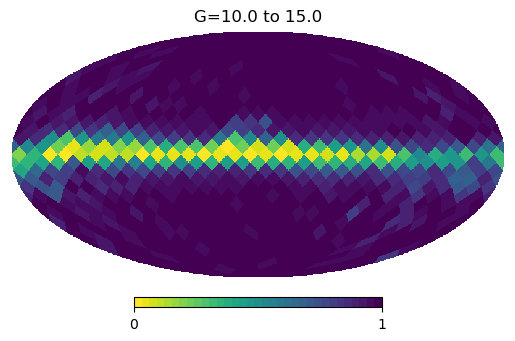}
\includegraphics[width=6.9 cm]{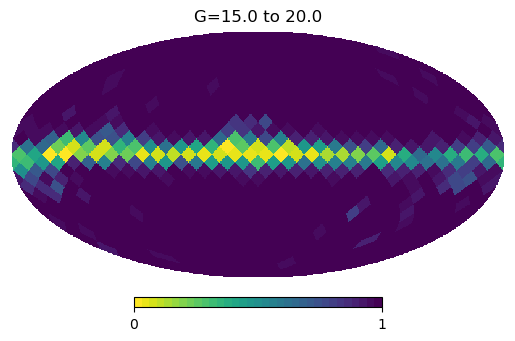}}
\caption{Completeness maps in galactic coordinates for the $S_{flag}$ in two ranges of apparent G~\mbox{magnitude}.\label{fig4}}
\end{figure}

Figure~\ref{fig4} shows the completeness maps of the selection function $S_{flag}$ for two ranges of apparent G magnitude. In~the areas close to the disc plane, completeness decreases to almost zero values, while outside the midplane of the disc it is close to $0.9$--
$1$  (or ~$90$--$100\%$). 

\unskip   
 
%Thus, the final selection function is the product of %the selection functions from the %equations~\ref{eq:scomb} and~\ref{eq:Cflag}. 

\subsection{Distance~Estimation}\label{sec:dist}
\begin{figure}[H]
%\isPreprints{\centering}{} % Only used for preprints
\includegraphics[width=12 cm]{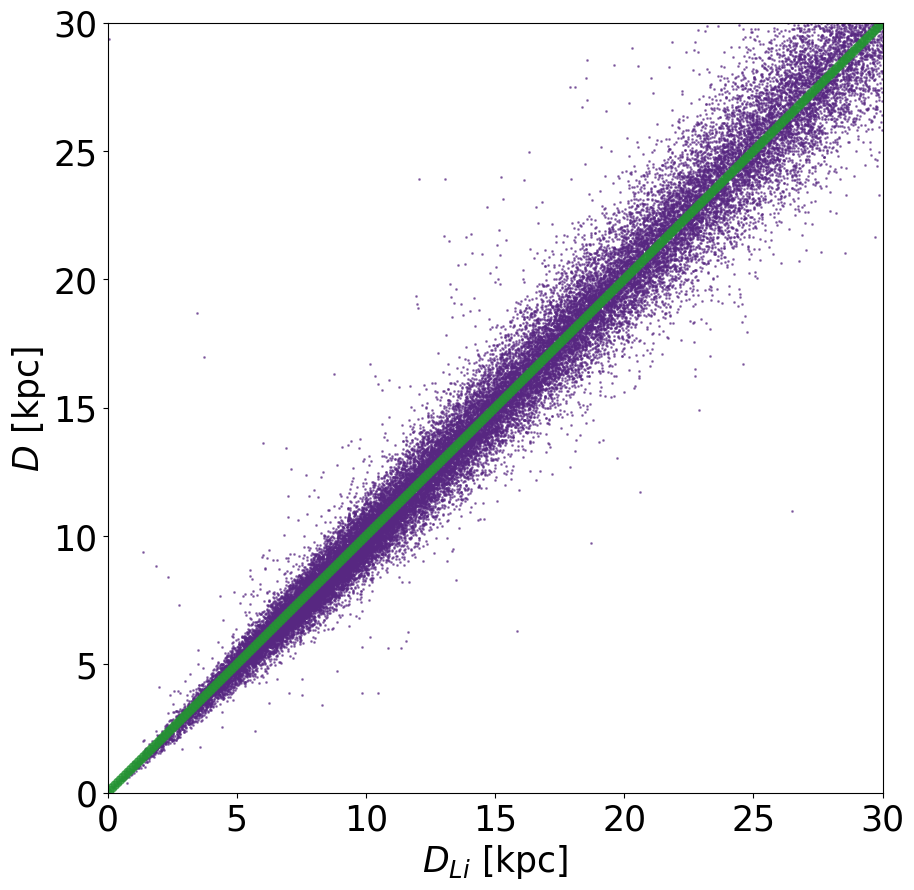}
\caption{Accurate distances from \citet{Li2023} versus distances estimated by us. The~green line indicates equal distances between both~samples.\label{Distance}}
\end{figure}

To determine the scale length and the scale height of the thick disc, we need to know a selection function not as a function of apparent G magnitude but~as a function of distance.
Only a fraction of stars (115,410 out of 183,755) have distances determined by \citet{Li2023}. 
%determined distances for which Fourier decomposition %parameters are available from \citet{Li2023}. 
For the stars lacking distances, we calculate them using the following equation:
\begin{equation}\label{eq:dist}
    G = M_G+A_G+5\log_{10}(D) - 5
    \,\textrm{.}
\end{equation}
where $G$ is the apparent magnitude of the star in the Gaia G-band, $M_{G}$ is its absolute magnitude, $D$ is its  distance in parsecs, and~$A_{G}$ is its extinction value. We take $M_{G}=0.6$, a~mean value determined by~\cite{Mat2020, Li2023}, as the absolute stellar magnitudes for RRab Lyrae stars. Following~\cite{Li2023}, for~stars with $|b|>25^{\circ}$, extinction was calculated using the expression from  $A_{G}=R_{G}~E(B-V)$, where $R_{G}=2.516$ \cite{Huang}, and~$E(B-V)$ taken from the reddening maps from~\cite{Shleg98}, taking into account the 14\% %please check meaning retained
systematic correction according to~\cite{Sch,Yuan}. For~the stars with latitudes $|b|\le25^{\circ}$, we take the extinction $A_{G}$ from~\cite{clemen}. For~the sample of stars with accurately determined distances from \citet{Li2023}, 
Figure~\ref{Distance} compares them with our estimated distances computed using Equation~(\ref{eq:dist}),
showing good~agreement. 

\begin{figure}[H]
\scalebox{.99}[0.99]{\includegraphics[width=6.9cm]{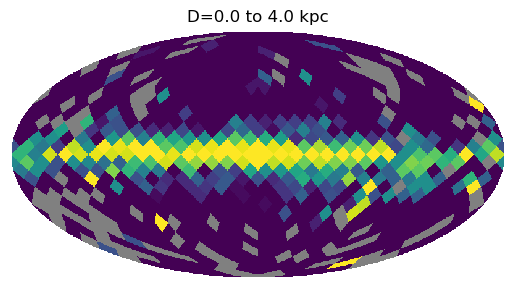}}
\scalebox{.99}[0.99]{\includegraphics[width=6.9cm]{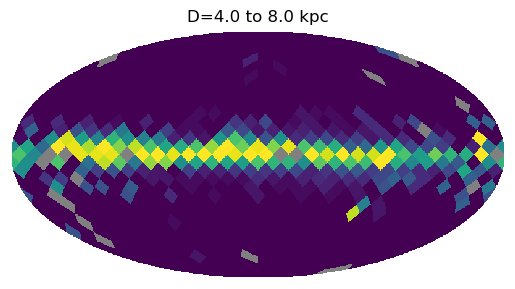}}
\scalebox{.99}[0.99]{\includegraphics[width=6.9cm]{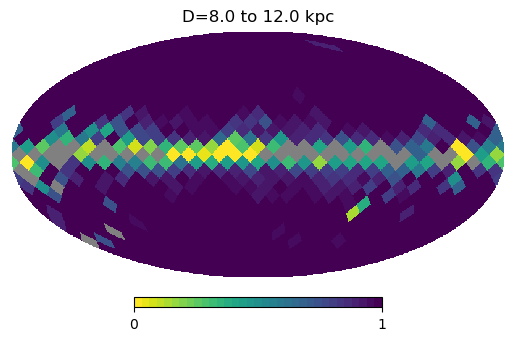}}
\scalebox{.99}[0.99]{\includegraphics[width=6.9cm]{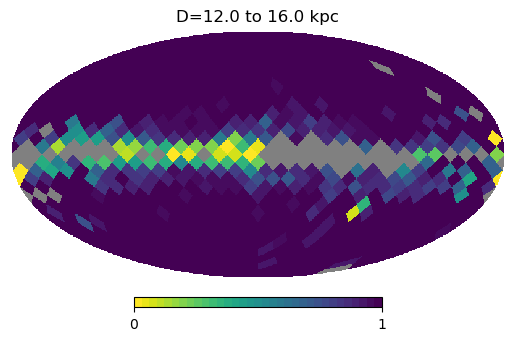}}
\caption{Completeness maps in galactic coordinates of the final selection function $S$ from Equation~(\ref{e:SA})
for four distance ranges. The~grey cells indicate the pixels where the stars do not fall.
\label{fig6}}
\end{figure}

To obtain the selection function $S$ from Equation~(\ref{e:Cdef}) as a function of distance, we compute the 
aforementioned completeness for each HEALpix and a certain distance interval. Figure~\ref{fig6} plots the final  selection function $S$ for four distance~ranges.

The next step in cleaning the sample is to eliminate regions related to globular clusters (GCs) and dwarf galaxies (DGs). To~eliminate GC %please check meaning retained
 RRLs, we remove all the stars within a 10~half-light radius of each globular cluster listed in the \citet{Harris} catalog of GCs. Following \citet{Garcia}, we exclude RRLs belonging to the Sagittarius Stream and Sagittarius Core by cross-matching their Gaia DR3 \texttt{source\_id} with the catalog of \citet{Ramos},
and the DGs were deleted using Table C.2 of \citet{GaiaDG}. We also removed all RRLs towards the Magellanic Clouds within a diameter of
 $16^{\circ}$ and  $12^{\circ}$ for Large and Small Clouds, respectively~\cite{Garcia}.
As mentioned above, we exclude the area with latitude of |b| < $10^{\circ}$ to avoid regions with a high content of dust and contaminations. We also excluded RRLs with a distance of $D>20$ kpc from our~consideration.

After all of the above data cleaning (quality and spatial cuts), our observational sample consists of 40,188 unique RRLs. Of~this final sample, 36,685 are listed in Gaia SOS and 33,703 in ASAS+PS1. There are 30,200 RRLs in common between both catalogs;~2,982 RRLs are only in Gaia SOS, and reciprocally, 3,503 are listed only in ASAS+PS1. We have 31,978 out of 40,188 RRLs with accurate distances from \citet{Li2023} and 8210 (20\%) with distances estimated by~us.

\unskip

%%%%%%%%%%%%%%%%%%%%%%%%%%%%%%%%%%%%%%%%%%
\section{Thick Disc and Halo Fitting~Model}\label{sec:model}
\unskip

\subsection{Density Profile~Models }\label{sec:density}

We assume that the density profile of the selected RRLs can be described in galactocentric cylindrical coordinates $(R,z)$ as a sum of the halo and thick disc distributions, assuming the absence or a negligible amount of thin disc RRLs (\citet{Minniti} have shown the absence of metal-rich RRLs, which can be associated with the
thin disc~\cite{MV}) in the Milky Way:
\begin{equation}\label{e:discc} 
C(R,z) =C_{disc}(R,z)+C_{Halo}(R,z). 
\end{equation}
%supposing that our sample is described by the thick %disc and the halo and there is no RRLs in the thin %disc. 

\subsubsection{The Thick~Disc }\label{sec:densitythick}

To model the thick disc density profile, we use two types of density distribution.
In the first one, we use the standard double exponential density profile \citep{MV,Juric}:
\begin{equation}\label{e:disc1} 
C_{Disc}(R,z) =C_{Disc,\odot} ~e^{-\tfrac{R-R_\odot}{h_R}} e^{-\tfrac{|z|}{h_z}}~. 
\end{equation}
where $h_z$, $h_R$ and $C_{Disc,\odot}$ are the vertical scale height,  the~radial scale length and the~concentration of RRLs in the solar neighborhood, respectively. $R_\odot$ is the distance of the Sun from the galactic center, taken to be 8.122 kpc. In~the second case, for the vertical density distribution of the thick disc, we assume the profile $sech^2$~\cite{GR,Juric,BT}:
\begin{equation}\label{e:disc2} 
C_{Disc}(R,z) =C_{Disc,\odot} ~e^{-\tfrac{R-R_\odot}{h_R}} sech^2(-z/{h_z}), 
\end{equation}
where $h_z$, $h_R$ and $C_{\odot}$ are the same values as in Equation~(\ref{e:disc1}). We do not consider the thick disc flare in our study because the thick disc is flat in the inner regions with radii of $R<14$--$16$ kpc \citep{Ham,R14}.  

\subsubsection{The~Halo }\label{sec:densityhalo}

For the halo density distribution, we use the standard power law~\cite{MV}:
\begin{equation} \label{e:rho_halo} 
 C_{Halo} (R,z) = C_{Halo,\odot} \left[ R^2 + \left(\frac{z}{q}\right)^2 \right]^{n/2} ,
\end{equation}
where $C_{Halo,\odot}=C_{Halo}/{R_\odot^n}$ is the halo RRL concentration in the solar vicinity, and~$q$ is the halo~flattening. 

From the kinematic parameters of stars from SDSS, \citet{Carollo} suggested a dual halo
model for the Milky Way galaxy, where the inner halo is dominated by the radial merger of some massive metal-rich clumps, while the outer halo consists of dissipation-less chaotic merging of smaller subsystems~\cite{liu}. Different breaking spherical radii ranging from 20 to 30 kpc~\cite{liu, Iorio, wat, Carollo} are proposed in articles where the parameter $n$ may vary due to the two-component halo. Since we are aiming to estimate the scale length and scale height of the thick disc, it is reasonable to limit ourselves to the inner halo with spherical radii of $r<20$ kpc. Excluding the region with $r<4$ kpc is necessary to avoid contamination from the bar/bulge subsystems (the bar extends up to $\sim$$5$~kpc, where its scale height is $\sim$$0.180$~kpc~\cite{weggb}).

\subsection{Fitting~Procedure}\label{sec:fit}

The fitting task is performed in two parts. First, we fit just the halo and find the parameters $n$ and $q$ (two free parameters fitting). For~this part, we restrict the sampled volume by $|z|>5$ kpc and $4<r<20$ kpc to avoid the outer halo, disc and central bulge/bar, and~we assume that the disc concentration $C_{disc}(R,z)$ in Equation~(\ref{e:discc}) is zero. The~number of stars in the experiment
is 10,404.

After that, we keep the parameters of the halo fixed and adjust the parameters of the thick disc $h_z$, $h_R$ and the ratio of the concentration of halo stars to the concentration of the thick disc $\gamma=C_{Halo,\odot}(R,z)/C_{Disc,\odot}(R,z)$ in the solar neighborhood (three free parameters fitting). When %please check meaning retained
 fitting the thick disc, we restrict our region by a cylindrical radius of $4<R<20$ kpc and a height of  $|z|<5 $ kpc. The~number of stars in the experiment
is 11,707. 

\begin{figure}[H]
%\isPreprints{\centering}{} % Only used for preprints
\scalebox{.99}[0.99]{\includegraphics[width=6.9 cm]{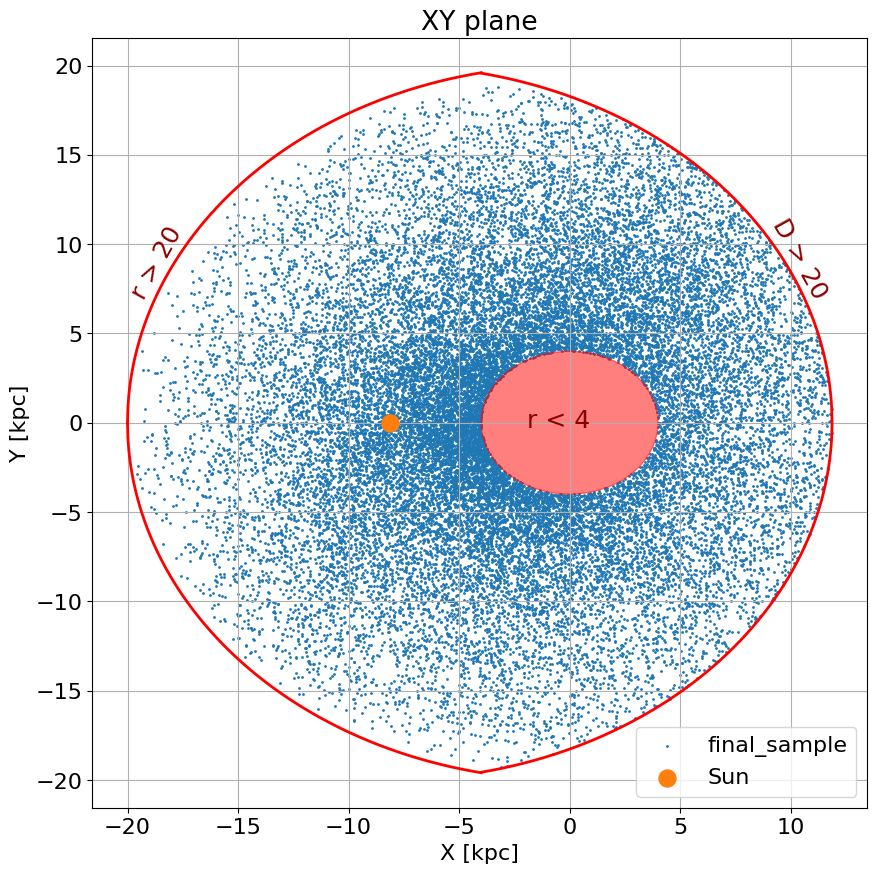}}
\scalebox{.99}[0.99]{\includegraphics[width=6.9 cm]{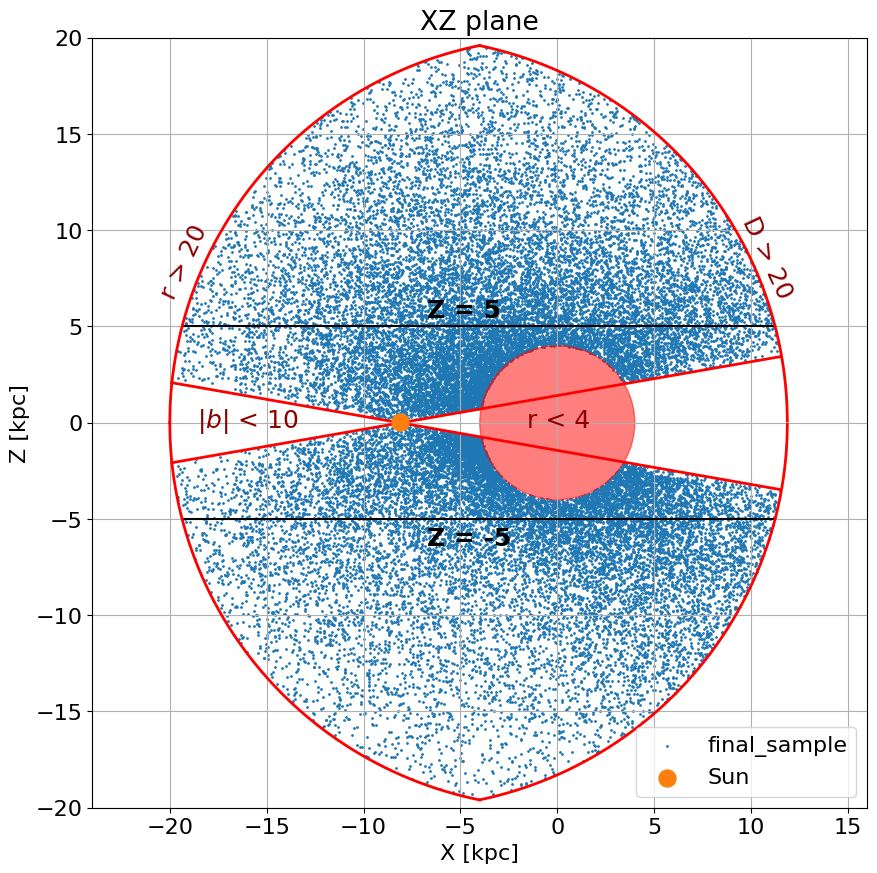}}
\caption{The $(x,y)$ and $(x,z)$ distribution of our final RRL sample in the galactocentric~coordinates.\label{xyz}}
\end{figure}

Figure~\ref{xyz} demonstrates distribution of our final RRL sample in the $(x,y)$ and $(x,z)$ plane, marking the areas where the fitting is performed. The~galactic center is located at $x=0$ kpc and $y=0$ kpc.

\begin{figure}[H]
%\isPreprints{\centering}{} % Only used for preprints
\includegraphics[width=12 cm]{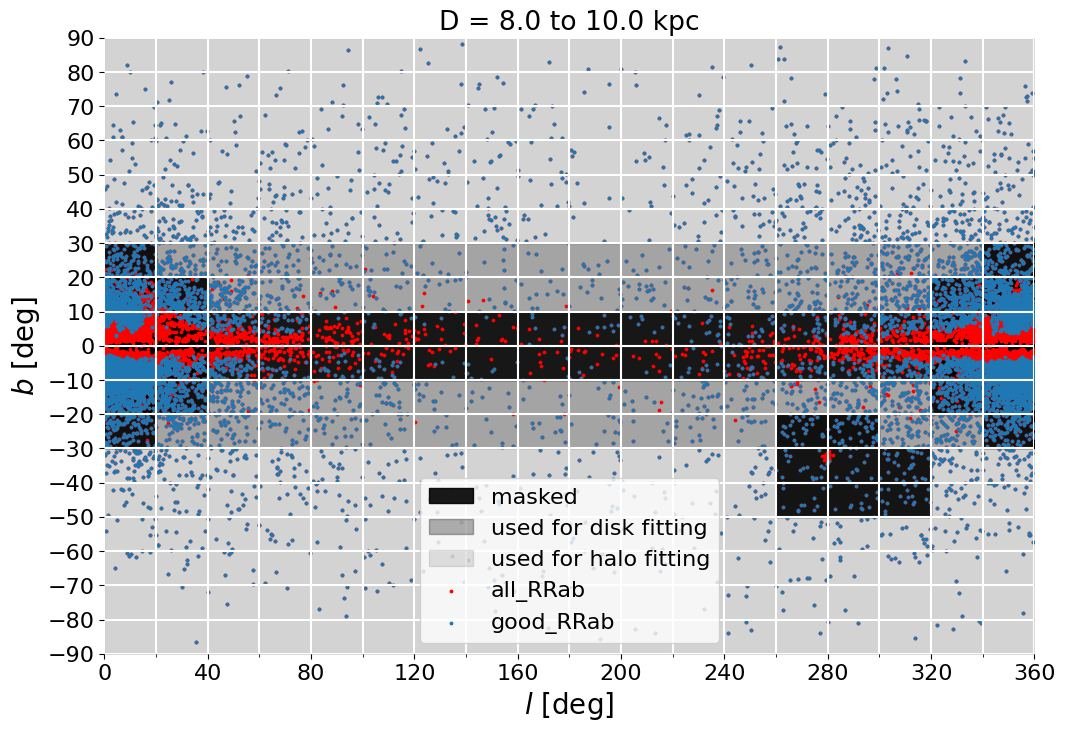}
\caption{The custom bins in the galactic coordinates for the slice of radii 
from 8 to 10 kpc with blue points (`good’ stars) and red points (`bad’ stars). The~bins with black backgrounds denote masked bins which are not used in the fitting. The~grey and dark grey colors mark the areas where the halo and the thick disc are fitted.\label{cust}}
\end{figure}

We use the following procedure: We split our galaxy into custom bins in the galactic coordinates $(l,b,D)$ with sizes of $20^{\circ}$, $10^{\circ}$ and~$2$ kpc, respectively, and~find the number of RRLs in each bin $i$ according to Equation~(\ref{e:discc}) as follows:
\begin{equation} \label{e:Number} 
N_i = \int_{V_i} C(l,b,D)\,dV = \int_{l_i}^{l_{i+1}} \int_{b_i}^{b_{i+1}}\int_{D_i}^{D_{i+1}} C\cdot cos(b)\cdot D^2 \, dl\cdot db\cdot dD,
\end{equation}
where $C(l,b,d)$ is the number density as a function of the galactic coordinates. The~HEALPixes are constructed as equal area pixels, which implies that there are certain levels of grid resolution. Since we integrate the density function within the bin, we are not forced to use equal area bins; therefore, we use our custom bins with more suitable sizes for our sample.
We use masks to remove the bins that contain Magellanic Clouds. 
We remove stars in the direction of globular clusters and dwarf galaxies, as~described above; however, we do not mask the corresponding bins because the spatial size of the removed area is significantly smaller compared to the bin size (<1 percent of the bin area for all clusters and dwarf galaxies).
Figure~\ref{cust} demonstrates our custom bins in the galactic coordinates for the slice of radii from 8 to 10 kpc with blue points (`good' stars) and red points (`bad' stars). Bins with black backgrounds describe masked bins, which are not used in the fitting. The~grey and dark grey colors mark the area where the halo and the thick disc are~fitted.

Then, to~find the selected number of RRLs in each bin, we multiply by the value of the selection function from Equation~(\ref{e:Cdef}):
\begin{equation} \label{e:Number} 
N_{selected, i} = N_i \cdot S_i
\end{equation}

Then, we construct the probability density function of the RRLs to be observed from the selected number of RRLs by normalizing the sum to 1. Finally, we derive the number of RRLs generated in each bin as a probability density function multiplied by the total number of observed RRLs. In~other words, we distribute the total number of observed RRLs among the bins according to the probability density function:
\begin{equation} \label{e:Numbersel} 
N_{gen, i} = \frac{N_{selected,i}}{\sum_{i=1}^{M} N_{selected,i}} \cdot \sum_{i=1}^{M}N_{obs,i}
\end{equation}
%Poisson likelihood $p(\lambda)$ (n - generated, m - observed)\\
%\begin{equation} \label{e:pxl} 
%p(x|\lambda) = \frac{e^{-\lambda}\lambda^x}{x!}\\
%\end{equation}
where M is the total number of~bins.

We estimate the parameters of the models through Bayes inference using the procedure from \citet{Tremmel_2013}.
We assume that the observed number of RRLs ($m_i$) in our bins is drawn from an underlying Poisson distribution
\begin{equation} \label{e:pmil} 
p(m_i|\lambda_i) = \frac{e^{-\lambda_i}\lambda_i^{m_i}}{m_i!}
\end{equation}
where we are not interested in the Poisson mean $\lambda$, so we marginalize over it.
\begin{equation} \label{e:pmt} 
p(m|\theta) = \prod_i \int p(m_i|\lambda_i)p(\lambda_i|\theta)d\lambda_i 
\end{equation}
where $\theta$ is the set of parameters used for~fitting.

Assuming that the generated number of RRLs ($n_i$) is distributed with the same Poisson distribution
\begin{equation} \label{e:pnl} 
p(n_i|\lambda_i) = \frac{e^{-\lambda_i}\lambda_i^{n_i}}{n_i!} ,
\end{equation}
we can estimate $p(\lambda_i|\theta)$ by applying Bayes' rule
\begin{equation} \label{e:plt} 
p(\lambda_i|\theta)=p(\lambda_i|n_i(\theta))=\frac{p(n_i|\lambda_i)p(\lambda_i)}{p(n_i)}\\,
\end{equation}
where $\theta$---model parameters.
Using Jeffreys prior on $\lambda_i$: $\displaystyle p(\lambda_i)= \frac{1}{\sqrt{\lambda_i}}$, we obtain
\begin{equation} \label{e:pplt} 
p(\lambda_i|\theta) = \frac{e^{-\lambda_i}\lambda_i^{n_i-\frac{1}{2}}}{\Gamma(n_i+\frac{1}{2})} \\
\end{equation}

Inserting Equations~(\ref{e:pmil}) and (\ref{e:pplt}) into Equation~(\ref{e:pmt}), we obtain
\begin{equation} \label{e:ppmt} 
p(m|\theta) = \prod_i \frac{\Gamma(m_i+n_i+\frac{1}{2})}{2^{m_i+n_i+\frac{1}{2}}\Gamma(n_i+\frac{1}{2})m_i!}
\end{equation}

Taking the logarithm and simplifying  
%
%\begin{equation} \label{e:logp} 
%    \log(p) = \sum_i^N \left[\log\Gamma(m_i+n_i+\frac{1}{2})
%        - (m_i+n_i+\frac{1}{2})\log2 -\log m_i!
%        - \log \Gamma(n_i+\frac{1}{2}) \right]
%\end{equation}
%
%
%
\begin{equation} \label{e:logpp} 
    \log(p) = \sum_i^M \left[\log\Gamma(m_i+n_i+\frac{1}{2})-
        \log \Gamma(n_i+\frac{1}{2}) \right]
        - \log2  (M+N+\frac{1}{2}) - \sum_i^M \log m_i!
\end{equation}
        
%N number of pixels and N total number of generated points. Change something!!!
And then omitting the constants
\begin{equation} \label{e:logppp} 
\log(p)\sim \sum_i^M \left[\log\Gamma(m_i+n_i+\frac{1}{2})-
        \log \Gamma(n_i+\frac{1}{2}) \right]
\end{equation}

Thus, when $m_i=n_i$, the likelihood $\log(p)$ reaches its maximum. Using uniform priors for parameters, we find that the posterior is 
\begin{equation} \label{e:pnew} 
p(\theta|m) = \frac{p(m|\theta)p(\theta)}{p(m)} \propto p(m|\theta)
\end{equation}

We used a Markov chain Monte Carlo (MCMC) method, specifically %please check meaning retained. Altenatively, consider "particularly" or "especially"
 the~sampler from the \texttt{emcee} \cite{Fore} Python package, to explore the posterior PDF. We used 16 chains per parameter, performing 3000 steps in total. After~convergence was reached (500 steps for the burn-in phase), we obtained the~results. 
 
As an additional test, we independently used the \texttt{Nautilus} \cite{nautilus} Python package, which uses %please check meaning retained
 the importance nested sampling (INS) technique for the Bayesian posterior, and~found essentially identical~results.

%%%%%%%%%%%%%%%%%%%%%%%%%%%%%%%%%%%%%%%%%%
\section{Results}\label{sec:res}
\begin{figure}[H]
\scalebox{1.1}[1.1]{\includegraphics[width=10.9cm]{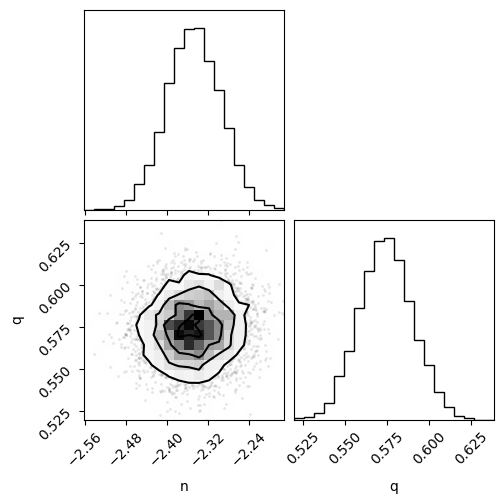}}
\caption{Two-dimensional histograms showing the distribution of values found in the MCMC simulation for the halo parameters n and flattening~q.\label{fig7}}
\end{figure}

All the best-fit parameter results in this study are presented as the median value and $\pm 1\sigma$ equivalent range, with the~latter being defined by the 15.87 and 84.13 percentiles. In~the first step of the fitting procedure, we determined the exponent of the halo power law density profile to be $n =-2.35_{-0.05}^{+0.05}$ and flattening $q =0.57_{-0.02}^{+0.02}$. Table~\ref{tab1} shows the obtained results and also includes the mean values and standard deviations. In~Figure~\ref{fig7}, we show the corresponding `corner plot', i.e.,~the density plot of the two-dimensional projections of the PDF, showing the relationship between the parameters $n$ and $q$. The~black circular lines correspond to areas with $ 1\sigma, 2\sigma, 3\sigma$ equivalent~ranges.

\begin{table}[H] 
%\tablesize{\small}
\caption{Best fitting parameters of the thick disc and~halo.\label{tab1}}
%\isPreprints{\centering}{} % Only used for preprints
\begin{tabularx}{\textwidth}{CCC}

\toprule
\multicolumn{3}{c}{\textbf{Halo}}  \\
\midrule
&\textbf{n}	& \textbf{q}	\\
\midrule
median $\pm$ range
 & $-2.35_{-0.05}^{+0.05}$		& $0.57_{-0.02}^{+0.02}$				\\
mean $\pm~1\sigma$  & $-2.35_{-0.05}^{+0.05}$		& $0.57_{-0.02}^{+0.02}$	\\
\end{tabularx}
\begin{tabularx}{\textwidth}{CCCC}
\midrule
\multicolumn{4}{c}{\textbf{Thick disc with exponential vertical profile} 
}  \\
\midrule
& \textbf{$h_R$}	& \textbf{$h_z$}	& \textbf{$\gamma$} \\
\midrule
median $\pm$ range & $-2.14_{-0.17}^{+0.19}$~kpc		& $0.64_{-0.06}^{+0.06}$~kpc	& $0.19_{-0.02}^{+0.02}$			\\
mean $\pm~1\sigma$  & $-2.14_{-0.18}^{+0.18}$~kpc		& $0.64_{-0.06}^{+0.06}$~kpc	& $0.19_{-0.02}^{+0.02}$			\\
\end{tabularx}
\begin{tabularx}{\textwidth}{CCCC}
\midrule
\multicolumn{4}{c}{\textbf{Thick disc with \boldmath{$sech^2$} vertical profile}}  \\
\midrule
& \textbf{$h_R$}	& \textbf{$h_z$}	& \textbf{$\gamma$} \\
\midrule
median $\pm$~range & $-2.10_{-0.17}^{+0.19}$~kpc		& $1.02_{-0.08}^{+0.09}$~kpc	& $0.32_{-0.03}^{+0.03}$			\\
mean $\pm~1\sigma$  & $-2.11_{-0.18}^{+0.18}$~kpc		& $1.02_{-0.08}^{+0.08}$~kpc	& $0.32_{-0.03}^{+0.03}$			\\
\bottomrule
\end{tabularx}
\noindent{\footnotesize{\textsuperscript{} 
 The halo power law density exponent $n$ and flattening $q$, the~thick disc scale length $h_R$ and height $h_z$, and the halo to disc concentration ratio $\gamma$ obtained in this work.}}
\end{table}

It is important to note that the resulting flattened inner halo of the RRLs should not be associated with the dark matter halo. The~explanation of the stellar halo's flatness is linked to the accretion history of the Milky Way and can be related, for~example, to~the dominant influence of the Gaia Sausage/Enceladus accretion at redshift $z \approx2$ \citep{TKG,NaiduG,BelokurovG}.

\begin{figure}[H]
\scalebox{1.1}[1.1]{\includegraphics[width=10.5cm]{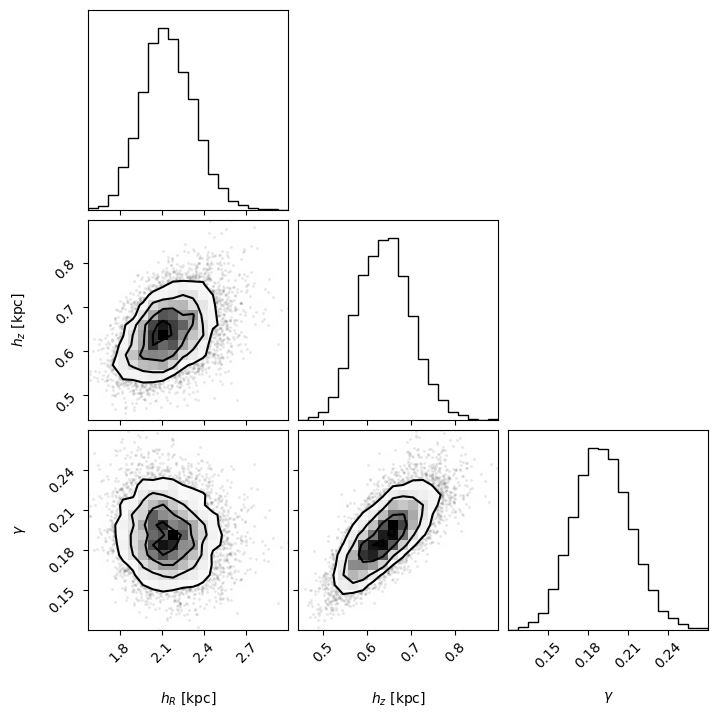}}
\caption{Two-dimensional histograms showing the distribution of values found in the MCMC simulation of the thick disc's parameters---$h_R$ and $h_z$---%please check meaning retained
	with the exponential vertical profile and halo to disc concentration $\gamma$.\label{fig8}}
\end{figure}

\begin{figure}[H]
\scalebox{1.1}[1.1]{\includegraphics[width=10.5cm]{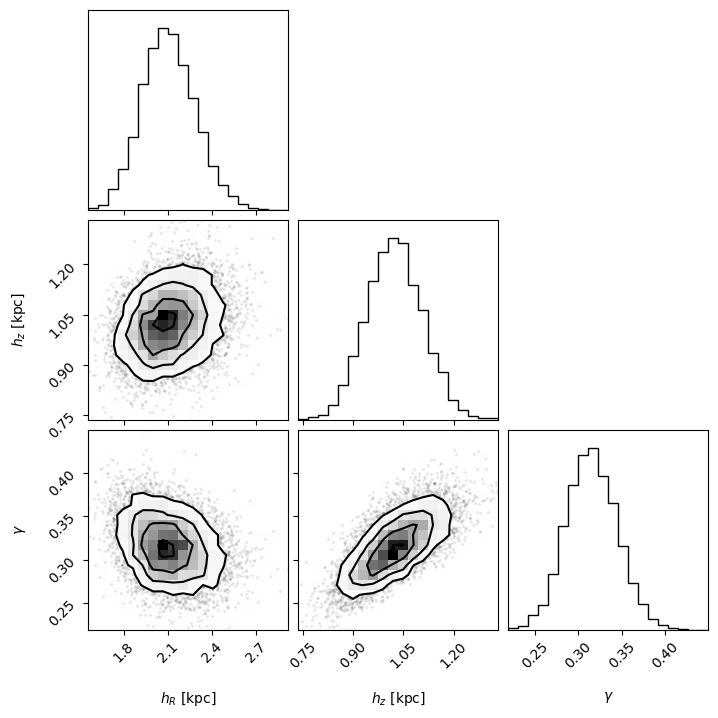}}
\caption{Two-dimensional histograms showing the distribution of values found in the MCMC simulation of the thick disc's parameters: $h_R$ and $h_z$ with $sech^2$ vertical profile and halo to disc concentration $\gamma$.\label{fig9}}
\end{figure}

At the second step, with~the halo parameters found above fixed, we fitted the parameters of the thick disc $h_z$, $h_R$ and the halo to the thick disc concentration ratio $\gamma=C_{Halo,\odot}(R,z)/C_{Disc,\odot}(R,z)$ in the solar neighbourhood for~each of the two disc vertical profile models: exponential and $sech^2$.
For the exponential profile, we found $h_R=2.14_{-0.17}^{+0.19}$ kpc, $h_z=0.64_{-0.06}^{+0.06}$ kpc and $\gamma=0.19_{-0.02}^{+0.02}$. For~the  $sech^2$ profile, we found $h_R=2.10_{0.17}^{+0.19}$ kpc, $h_z=1.02_{-0.08}^{+0.09}$ kpc and $\gamma=0.32_{-0.03}^{+0.03}$.
Table~\ref{tab1} also shows the best-fit parameters of the thick disc and halo to disc concentration ratio for both vertical profile models. \mbox{Figures~\ref{fig8} and~\ref{fig9}} show the corresponding corner plots. For~both models, we obtain almost identical thick disc radial scales; however, the~choice between the two vertical scale laws has a noticeable effect on determining the ratio of the halo to the disc concentration from $19\%$ to $32\%$ for the exponential and $sech^2$ vertical profile,~respectively.

%\unskip  

We also performed the five-parameter fitting of the halo and thick disc simultaneously, limiting the analysis to the spherical radii of $4<r<20$ kpc. We obtained very similar results, with the vertical scale showing the greatest deviation in the case of the $sech^2$ vertical profile, reaching a 10$\%$ difference. However, taking into account the errors, we obtained the same results with both approaches. These results can be found in Appendix~\ref{app}. 

\section{Discussion}\label{sec:discuss}

The parameters of the thick disc are determined by the mechanism of its formation. A~few scenarios for explanation of the origin of the thick disc have been suggested, which can be divided into three groups. The~first type of scenario explains the origin of the
thick disc by heating of the pre-existing thin disc. The~first physical mechanism for such
heating was suggested by \citet{spt}, who proposed that
the scattering of thin disc stars by molecular clouds could produce the thick disc.
Later, other mechanisms were proposed, including heating by spiral structure (\citet{SC}), minor mergers with accreted small satellites~\cite{Quinn},
and radial migration of stars, as considered by \citet{ShB}, as possible mechanisms for thick disc formation%please check meaning retained
.
Another approach to explain the origin of the Milky Way's thick disc involves a major accretion of a satellite galaxy. \citet{ABADI} suggests that the thick disc was
formed from the stars born in a disrupted massive satellite galaxy. An~“intermediate”
scenario of thick disc formation assumes that gas-rich mergers trigger an
episode of thick disc formation. \citet{GRAND} analyzed the Auriga
cosmological zoomed%please check meaning retained
-in Milky Way-like galaxy simulations and came to the conclusion that the formation of the thick disc was caused by a massive Gaia Sausage/Enceladus-like merger with a massive %please check meaning retained
gas-rich satellite with a mass of 10$^{8-9.5}$ $M_{\odot}$, which caused a dual effect on the evolution of the Milky Way galaxy. It “splashed” the existing
proto-disc stars into the halo, brought fresh gas into the central regions of the galaxy
and triggered a starburst that formed a thick~disc.

\citet{park} analyzed the results of high-resolution cosmological simulations in GALACTICA and NEWHORIZON to understand whether the spatially distinct thin and
thick discs are formed by different mechanisms. These authors traced the birthplaces of the stellar particles in the thin and thick discs and found that most of the
thick disc stars in simulated galaxies were formed close to the midplane of their discs, which suggests that the two discs are not distinct in terms of formation mechanism but
rather are the signature of a complex formation of the Milky Way's disc. This scenario is supported by the findings of \citet{V22}, who used RGB stars from the GAIA DR3
catalog and found that about half of the stars in the solar neighborhood have
thick disc kinematics. Later, \citet{Vieira} confirmed this result using a complete
sample of about 330,000 dwarf stars which were well-measured by Gaia DR3, finding that the ratio of the concentration of the thick disc dwarf stars to the concentration of stars with thin disc kinematics in the solar neighborhood is 0.75 $\pm$ 0.05, which is consistent with the result based on the RGB-stars. Finding that the thick stellar disc is as massive as the
thin one changes our understanding of thick disc formation. If~the mass of the
Milky Way's thick disc is comparable to that of the thin one, i.e.,~in the order of 10$^{10}$ solar masses, then an accreted gas-rich massive satellite galaxy should have a mass of about a
few times 10$^{10}$--10$^{11}$ solar masses. The accretion of such massive satellites will destroy a pre-existing disc, and~the result of such accretion could lead to the formation of an elliptical galaxy. The~formation of the thick disc looks rather like an evolutionary stage in the formation of the Milky Way's multi-component and multi-structured thick disc. In~light of this, we can understand the discrepancies in estimates of the radial scale length and the scale height of the thick disc based on the different objects mentioned above. It might be that the oldest thick disc objects, like RR Lyrae or BHB stars, trace the early stages of building the modern Milky Way's thick disc, while the main sequence stars trace its later~stages.

As mentioned in the introduction section, several attempts have been made to determine the parameters of the thick disc of the Milky Way. For~example, using UBV color--magnitude diagram fitting, \citet{R96} found the galactic thick disc radial scale length and scale height to be $2.8 \pm 0.8$ kpc and $0.76 \pm 0.05$ kpc, respectively. \citet{R14}, also using color--magnitude diagram (CMD) fitting, found those values to be $2.3$ kpc and $0.47$ kpc, respectively, for the $sech^2$ vertical profile. Using the photometric parallax for 48~million stars detected by the Sloan Digital Sky Survey (SDSS), \citet{Juric} determined values of $h_R=2.6$ kpc and $h_z=0.9$ (for the exponential profile) kpc, respectively, for the thick disc. \citet{Bensby11}, using K-giant stars, determined the thick scale length to be $2.0$~kpc. \citet{Bo11}, using G-dwarfs from SDSS, determined the thick disc $h_R$ and $h_z$ to be $2.01$ kpc and $0.686$ kpc (for the exponential profile), respectively. Recently, \citet{Vieira}, using a complete sample of about 330,000 dwarf stars that were well-measured by Gaia DR3, found the thick disc scale height to be $0.797 \pm 0.012$ kpc for the $sech^2$ vertical profile. \citet{MV} listed in Table~6 the scale height and the scale length of the thick disc obtained by different authors. As~one can see from this Table, the~radial scale length determinations vary from 1.9~$\pm$~0.1~\cite{MAC}
to 4.7~$\pm$~0.2~\cite{LH}. The~determinations of the vertical scale height of the thick disc vary from 0.51~$\pm$~0.04~\cite{Carollo10} to 1.06~$\pm$~0.05~\cite{cab}.

Using a combination of public RR Lyrae star catalogs totaling 1,305 stars and Bayesian fitting, \citet{MV} found $h_R=2.1$ kpc and $h_z=0.65$ kpc (for the exponential profile), respectively, using an exponential vertical profile law. Our results of $h_R=2.14 ^{+0.19}_{-0.17}$ kpc and $h_z=0.64\pm 0.06$ kpc for the same profile coincide very well with these findings. As~for the halo parameters, \citet{MV} determined $n =-2.78_{-0.05}^{+0.05}$ and $q =0.9_{-0.03}^{+0.05}$, which are different from ours. Finally, their halo to disc concentration ratio was estimated to be $23$--$27\%$, while our results extend over a slightly larger range, from 19--32\%. It is worth noting that our RRL sample is $\sim$$20$ times larger than~theirs.

Other investigations of the Milky Way's halo have relied, in many cases, on RRLs. For~example, \citet{Sesar13} found $n=-2.42$ and $q=0.63$ using RRLs, which are close to our values. Similarly, \citet{Sesar11} found $n=-2.62 \pm 0.04$ and $q=0.7 \pm 0.02$. Recently, \citet{IorioRR}, using a Gaia DR3 RRL sample, found a significant flattening in the Milky Way's stellar halo, with $0.5\lesssim q \lesssim 0.8$. Finally, using the photometric parallax for 48 million stars in SDSS, \citet{Juric} determined the values $n=-2.8 \pm 0.2$ and $q=0.64 \pm 0.1$. Our result of $n=-2.35\pm 0.05$ is slightly higher and indicates %please check meaning retained
flattening, which is consistent with their numbers. In~general, the~parameters $n$ and $q$ vary widely in the literature, with~$-3.75\lesssim n \lesssim -2.3$ and $0.5\lesssim q \lesssim 1$ \cite{Sesar11,Sesar13,MV,Juric,SM,Deason,IorioRR}. 

Overall, our investigation and results stand out for having used a very large sample (+20,000 stars) of high-quality (RRLs) data, giving robustness to the estimated~parameters.

%%%%%%%%%%%%%%%%%%%%%%%%%%%%%%%%%%%%%%%%%
\section{Conclusions}\label{sec:conclusion}

Our study of the galactic thick disc and internal stellar halo has provided insights into their density distributions by~using the most complete sample available of RRLs built from the Gaia DR3 SOS, ASAS-SN-II and PanSTARRS1~catalogs. 

We applied Bayesian methods to find the thick disc scale length and height as well as the halo parameters of the Milky Way. We used an MCMC approach to determine these parameters and their statistical uncertainties. We also employed the INS technique for the Bayesian posterior and found essentially identical~results.

We have determined the exponent of power law density profile of the halo to be $n =-2.35_{-0.05}^{+0.05}$ with flattening of $q =0.57_{-0.02}^{+0.02}$. The~thick disc scale length is $h_R=2.14_{-0.17}^{+0.19}$ kpc, the scale height is $h_z=0.64_{-0.06}^{+0.06}$ kpc and the concentration ratio (halo to disc) is $\gamma=0.19_{-0.02}^{+0.02}$ for the $sech^2$ vertical profile. For~the exponential vertical profile, the values are $h_R=2.10_{0.22}^{+0.19}$ kpc, $h_z=1.02_{-0.08}^{+0.09}$ kpc and $\gamma=0.32_{-0.03}^{+0.03}$.

Further progress in parameter refinement may be made with future Gaia releases, as well as progress in measuring the metallicities of RRLs and their radial velocities, which would provide complete 6D phase spaces for each star. This would allow us to more clearly separate the contributions of the halo and the thick disc RRL. In~addition, it would accurately answer the question of whether RRL stars are present in the thin disc. It is also worth noting that, in addition to existing surveys of RR Lyrae such as Gaia, ASAS and PS1, the upcoming %please check meaning retained
 Vera Rubin Observatory project---one of whose goals is mapping the structure %please check meaning retained
  of the Milky Way---will further expand our~dataset.

\vspace{6pt}

%%%%%%%%%%%%%%%%%%%%%%%%%%%%%%%%%%%%%%%%%%
\authorcontributions{Conceptualization, A.L., R.T., V.K., G.C., and K.V.; methodology, R.T., A.L., V.K., and G.C.; data curation, R.T. and A.L.; project administration, V.K.; supervision, V.K., G.C., and K.V.; validation, V.K., G.C., and K.V.; visualization, R.T. and A.L.;  
software, R.T. and A.L.; writing---original draft preparation, R.T and A.L.; writing---review and editing, K.V., V.K., R.T., A.L., and G.C. All authors have read and agreed to the
published version of the~manuscript.}

\funding{R.T. thanks the Foundation for the Advancement of Theoretical Physics and Mathematics `BASIS' for financial support:
 \url{https://basis-foundation.ru}.}

%\institutionalreview{Not applicable.}

%\informedconsent{Not applicable.}

\dataavailability{All data used in this paper were collected from open sources and the
references are provided.}

\acknowledgments{We thank the anonymous reviewers for their careful reading of the article and
valuable comments. R.T. thanks the Foundation for the Advancement of Theoretical Physics and Mathematics `BASIS' for financial support: \url{https://basis-foundation.ru}. V.K. thanks the Ministry of Science and the
Higher Education of the Russian Federation for support under %please check meaning retained
 state contract
no. GZ0110/23-10-IF.}

\conflictsofinterest{The authors declare no conflict of~interests.}

%%%%%%%%%%%%%%%%%%%%%%%%%%%%%%%%%%%%%%%%%%
%% Optional

%% Only for journal Encyclopedia
%\entrylink{The Link to this entry published on the encyclopedia platform.}
\clearpage 
\abbreviations{Abbreviations}{
The following abbreviations are used in this manuscript:\\

\noindent 
\begin{tabular}{@{}ll}
SDSS & Sloan Digital Sky Survey\\
RRLs & RR Lyrae stars\\
CMD & Color--magnitude diagrams \\
MCMC & Markov chain Monte Carlo\\
INS & Omportance nested sampling \\
SOS & Specific Objects Study \\
ASAS & ASAS-SN-II\\
PS1 & PanSTARRS1\\
\end{tabular}
}

%%%%%%%%%%%%%%%%%%%%%%%%%%%%%%%%%%%%%%%%%%
%% Optional
\appendixtitles{yes} % Leave argument "no" if all appendix headings stay EMPTY (then no dot is printed after "Appendix A"). If~the appendix sections contain a heading then change the argument to "yes".
\appendixstart
\appendix

\section[\appendixname~\thesection]{Five Parameter Fitting} \label{app}
\begin{figure}[H]
\includegraphics[width=13.2cm]{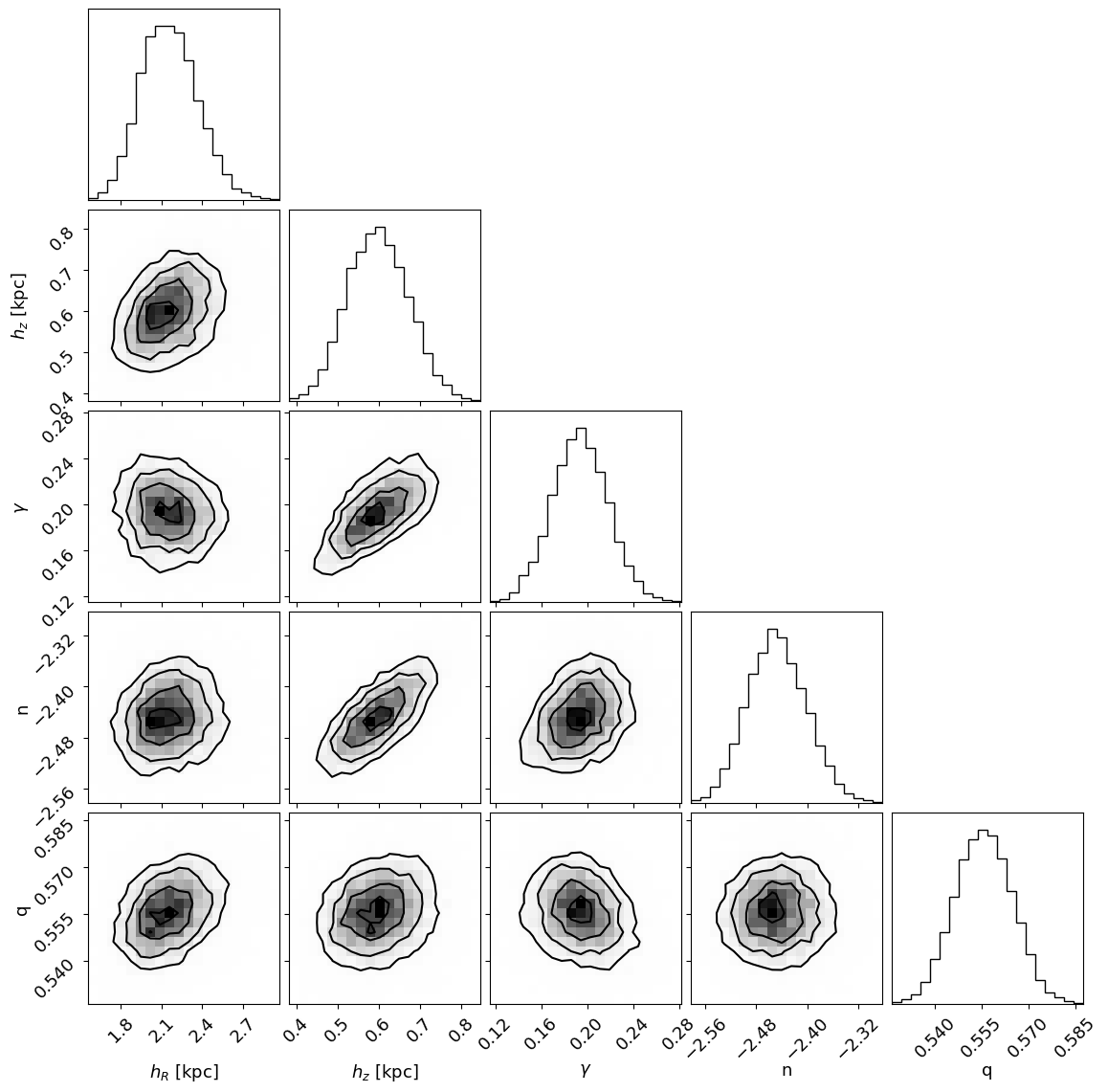}
\caption{Five-dimensional histograms showing the distribution of values found in the MCMC
simulation for the thick disc and halo parameters: $h_R$ and $h_z$ with exponential  profile, halo to disc
concentration $\gamma$, n and flattening q.\label{fig10}}
\end{figure}

\begin{figure}[H]
\includegraphics[width=13.2cm]{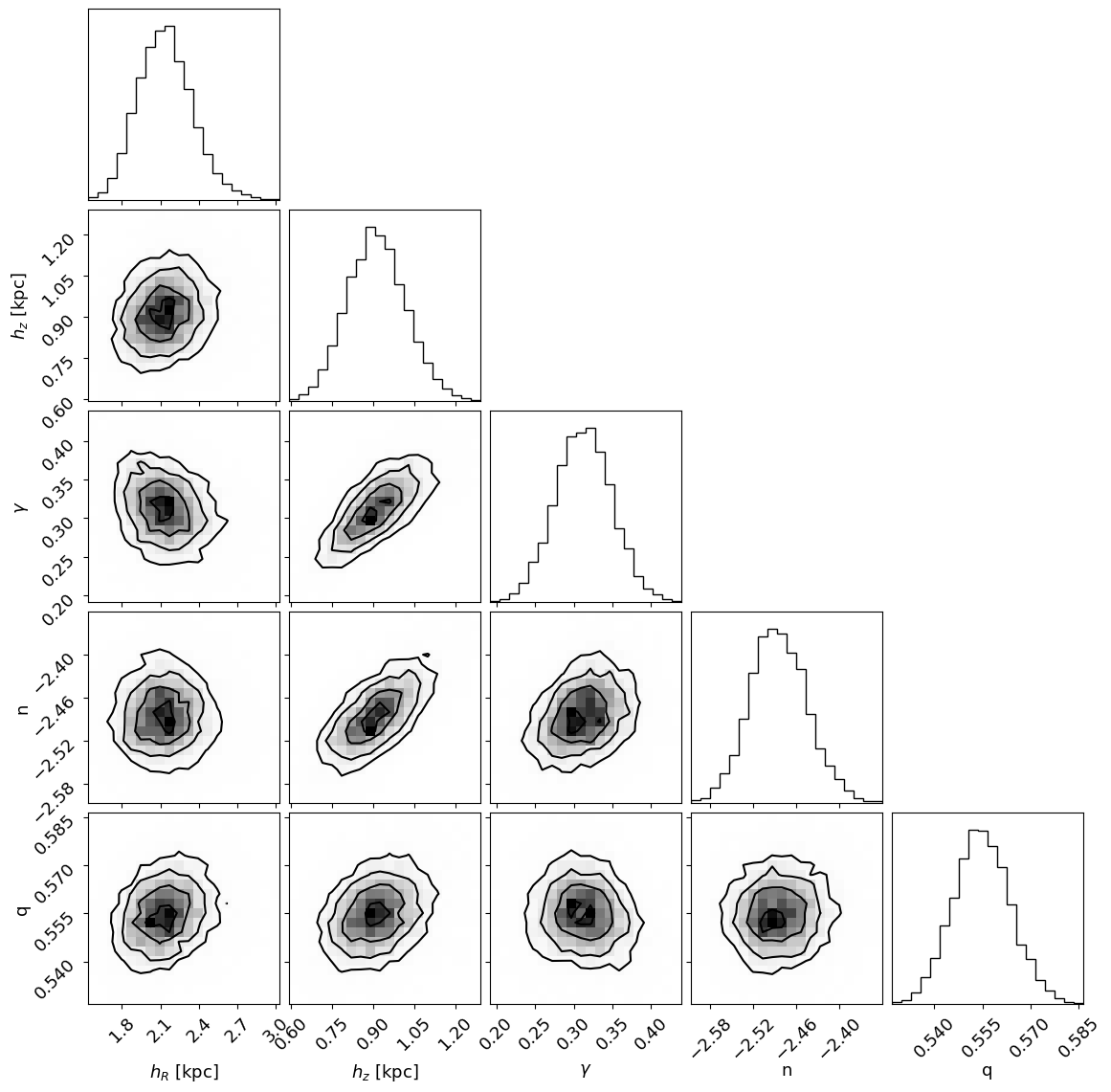}
\caption{Five-dimensional histograms showing the distribution of values found in the MCMC
simulation for the thick disc and halo parameters: $h_R$ and $h_z$ with $sech^2$ vertical profile, halo to disc
concentration $\gamma$, n and flattening q.\label{fig11}}
\end{figure}

%\subsection[\appendixname~\thesubsection]{}
As a test, we performed the simultaneous fitting of the halo and disc parameters without splitting the task onto two parts. In~this case, we limited the data to be contained between spherical radii $4<r<20$ kpc and latitude $|b|>10$. Table~\ref{tab1a} shows the best-fit parameters of the halo and the thick disc's parameters with median and mean values for both models of vertical disc profile: exponential and $sech^2$. Figures~\ref{fig10} and~\ref{fig11} show the corresponding corner plot. The~number of stars in the experiment
is 24,231.

%\unskip

\unskip

\begin{table}[H] 
%\tablesize{\small}
\caption{Best fitting 
 parameters of the thick disc and halo from five-parameters~fitting.\label{tab1a}}

\begin{tabularx}{\textwidth}{cCCCCC}
\toprule
\multicolumn{6}{c}{\textbf{Thick disc with exponential vertical profile and halo}}  \\
\toprule
& \boldmath{$h_R~[kpc]$ }	& \boldmath{$h_z~[kpc]$}	& \boldmath{$\gamma$}& \textbf{n} & \textbf{q} \\
\midrule
median $\pm$~range & $2.15_{-0.20}^{+0.22}$		& $0.60_{-0.07}^{+0.07}$	& $0.19_{-0.02}^{+0.02}$	&	$-2.45_{-0.05}^{+0.05}$ &	$0.56_{-0.01}^{+0.01}$ \\
mean $\pm~1\sigma$  & $2.16_{-0.21}^{+0.21}$		& $0.60_{-0.07}^{+0.07}$	& $0.19_{-0.02}^{+0.02}$	&	$-2.45_{-0.06}^{+0.06}$ &	$0.56_{-0.01}^{+0.01}$ 			\\
\end{tabularx}
\begin{tabularx}{\textwidth}{cCCCCC}
\toprule
\multicolumn{6}{c}{\textbf{Thick disc with \boldmath{$sech^2$} vertical profile and halo}}  \\
\toprule
&  \boldmath{$h_R~[kpc]$ }	& \boldmath{$h_z~[kpc]$}	& \boldmath{$\gamma$}& \textbf{n} & \textbf{q}  \\
\midrule
median $\pm$~ range
 & $2.13_{-0.20}^{+0.22}$		& $0.91_{-0.11}^{+0.11}$	& $0.31_{-0.04}^{+0.04}$	&	$-2.49_{-0.04}^{+0.04}$ &	$0.56_{-0.01}^{+0.01}$ \\
mean $\pm~1\sigma$  & $2.14_{-0.21}^{+0.21}$		& $0.91_{-0.11}^{+0.11}$	& $0.31_{-0.04}^{+0.04}$	&	$-2.49_{-0.04}^{+0.04}$ &	$0.55_{-0.01}^{+0.01}$ 				\\
\bottomrule
\end{tabularx}
\noindent{\footnotesize{\textsuperscript{} The exponent of power law halo density $n$, flattening $q$, thick disc scale length $h_R$ and height $h_z$ and halo to disc concentration ratio $\gamma$.}}
\end{table}

\begin{adjustwidth}{-\extralength}{0cm}

\setenotez{list-name=Note}

\printendnotes[custom]

\end{adjustwidth}

%%%%%%%%%%%%%%%%%%%%%%%%%%%%%%%%%%%%%%%%%%
%\isPreprints{}{% This command is only used for ``preprints''.
\begin{adjustwidth}{-\extralength}{0cm}
%} % If the paper is ``preprints'', please uncomment this parenthesis.
%\printendnotes[custom] % Un-comment to print a list of endnotes

\reftitle{References}
\PublishersNote{}
%\isPreprints{}{% This command is only used for ``preprints''.
\end{adjustwidth}
%} % If the paper is ``preprints'', please uncomment this parenthesis.
\end{document}